\documentclass[12pt]{article}
\usepackage{amssymb}
\usepackage{epsfig}
\usepackage{float}
\usepackage{graphicx}
\usepackage{amsmath}
\newcommand{\nua}[1]{\ensuremath{\rlap
           {\kern-2.5pt\ensuremath
           {\overset{\scriptscriptstyle(-)}{\phantom{\nu}}}}
           {\ensuremath{{\nu}_{#1}}}}}

\begin{document}
\begin{center}
{\bf On the phenomenology of neutrino oscillations in vacuum }

\end{center}
\begin{center}
S. M. Bilenky
\end{center}
\begin{center}
{\em  Joint Institute for Nuclear Research, Dubna, R-141980,
Russia\\}
{\em TRIUMF
4004 Wesbrook Mall,
Vancouver BC, V6T 2A3
Canada\\}
\end{center}
\begin{abstract}
A simple method of the calculation of neutrino transition probabilities in vacuum in the general case of $n$ massive neutrinos is presented. The method proposed  fully utilizes the unitarity of the mixing matrix. Three-neutrino case for both neutrino mass hierarchies is considered in some details. Transitions in the case of the sterile neutrinos are also discussed.
\end{abstract}
\section{Introduction}
The observation of neutrino oscillations in atmospheric
Super-Kamiokande \cite{SK}, solar SNO \cite{SNO}, reactor KamLAND \cite{Kamland} and other neutrino experiments \cite{Others} is one of the most important recent discovery in the particle physics.

Neutrino oscillations are based on the assumption that states of flavor neutrinos $\nu_{e}, \nu_{\mu}, \nu_{\tau}$ and sterile
neutrinos $\nu_{s_{1}}, \nu_{s_{2}},...$ are described by the coherent superpositions of  the states of  neutrinos with definite masses
(see ,for example, \cite{BilPont}, \cite{BPet}, \cite{BGG}, \cite{Conca},\cite{BilGiunti})
\begin{equation}\label{mixedstate}
|\nu_\alpha\rangle
=
\sum^{n}_{i=1} U_{\alpha i}^* \,~ |\nu_i\rangle.
\end{equation}
Here $n=3+n_{s}$ ($n_{s}$ is the number of sterile neutrinos) , $U$ is an unitary $n\times n$ mixing matrix, $|\nu_i\rangle$ is left-handed state of neutrino with mass $m_{i}$. The states $|\nu_\alpha\rangle$ satisfy the condition
\begin{equation}\label{mixedstate1}
    \langle\nu_{\alpha'}|\nu_\alpha\rangle=\delta_{\alpha'\alpha}.
\end{equation}
If at $t=0$ the flavor neutrino $\nu_\alpha$ is produced at the time $t$ we have
\begin{equation}\label{mixedstate2}
|\nu_\alpha\rangle_{t}= \sum_{\alpha'}|\nu_{\alpha'}\rangle\langle\nu_{\alpha'}|
e^{-iH_{0}t}|\nu_\alpha\rangle,
\end{equation}
where $H_{0}$ is the free Hamiltonian.

Thus, the probability of the transition $\nu_\alpha\to \nu_{\alpha'}$ during time $t$ is given by the expression
\begin{equation}\label{probability}
P( \nu_\alpha\to \nu_{\alpha'})= |\langle\nu_{\alpha'}|
e^{-iH_{0}t}|\nu_\alpha\rangle|^{2}=|\sum_{i}\langle\nu_{\alpha'}
|\nu_i\rangle e^{-iE_{i}t}\langle\nu_i|
\nu_\alpha\rangle|^{2}= |\sum_{i}U_{\alpha'i}
e^{-iE_{i}t}U^{*}_{\alpha i}|^{2}.
\end{equation}
It is obvious that the factor $ U^{*}_{\alpha i}$ is the amplitude of the transition from initial flavor state into the state with definite mass
$|\nu_i\rangle$, the factor $e^{-iE_{i}t}$ describes propagation in this state and the factor $ U_{\alpha' i}$ is the amplitude of the transition from the state with definite mass into the final state $|\nu_{\alpha'}\rangle$. For the ultrarelativistic neutrino we have
\begin{equation}\label{probability1}
 E_{i}=\sqrt{p^{2}+m^{2}_{i}}\simeq E+ \frac{m^{2}_{i}}{2E},
\end{equation}
where $E=p$ is the energy of neutrino at $m_{i}\to 0$.

The expression (\ref{probability}) can be rewritten in the form
\begin{equation}\label{probability2}
P( \nu_\alpha\to \nu_{\alpha'})=|\sum_{i}U_{\alpha'i}
e^{-2i\Delta _{pi}}U^{*}_{\alpha i}|^{2}.
\end{equation}
Here $p$ is an arbitrary fixed index,
\begin{equation}\label{probability3}
\Delta _{pi}=(E_{i}-E_{p})t\simeq \frac{\Delta m^{2}_{pi}L}{4E},\quad
\Delta m^{2}_{pi}=m^{2}_{i}-m^{2}_{p}.
\end{equation}
We used the relation
\begin{equation}\label{probability4}
t=L,
\end{equation}
where $L$ is the source-detector distance\footnote{Let us notice that
the validity of the relation (\ref{probability4}) was confirmed in the high-accuracy recent OPERA measurement \cite{Opera}}

From (\ref{probability2}) it obvious that transitions between different neutrinos are possible if two conditions are satisfied
\begin{enumerate}
  \item At least one neutrino mass-squared difference
 is different from zero.
  \item Neutrinos are mixed ($U\neq 1$).
\end{enumerate}

\section{Standard expression for the transition probability}

From (\ref{probability2}) we obviously have
\begin{equation}\label{standard}
P( \nu_\alpha\to \nu_{\alpha'})=\sum_{i}|U_{\alpha' i}|^{2}  |U_{\alpha i}|^{2}+ 2~\mathrm{Re}\sum_{i>k}U_{\alpha'i}
U^{*}_{\alpha i}U^{*}_{\alpha'k}
U_{\alpha k}  e^{-2i\Delta _{ki}}.
\end{equation}
From the unitarity  of the matrix $U$ ($\sum_{i}U_{\alpha' i}~U^{*}_{\alpha i}=\delta_{\alpha' \alpha}$) we find
\begin{equation}\label{standard1}
 \sum_{i}|U_{\alpha' i}|^{2}  |U_{\alpha i}|^{2}+ 2~\mathrm{Re}\sum_{i>k}U_{\alpha'i}
U^{*}_{\alpha i}U^{*}_{\alpha'k}
U_{\alpha k}  =\delta _{\alpha' \alpha}
\end{equation}
From (\ref{standard}) and (\ref{standard1}) for the transition probability we obtain the following expression 
\begin{equation}\label{standard2}
P( \nu_\alpha\to \nu_{\alpha'})=\delta _{\alpha' \alpha}
-2~\mathrm{Re}\sum_{i>k}U_{\alpha'i}
U^{*}_{\alpha i}U^{*}_{\alpha'k}
U_{\alpha k}(1-  e^{-2i\Delta _{ki}}).
\end{equation}
Finally we obtain the following standard expression for the probability of neutrino transition in vacuum (see \cite{Conca}, \cite{Giunti})
\begin{equation}\label{standard3}
P( \nu_\alpha\to \nu_{\alpha'})=\delta _{\alpha' \alpha}
-4~\sum_{i>k}\mathrm{Re}U_{\alpha'i}
U^{*}_{\alpha i}U^{*}_{\alpha'k}
U_{\alpha k}\sin^{2}\Delta _{ki}+2~\sum_{i>k}\mathrm{Im}U_{\alpha'i}
U^{*}_{\alpha i}U^{*}_{\alpha'k}
U_{\alpha k}\sin 2\Delta _{ki}
\end{equation}
In order to obtain $\bar\nu_\alpha\to \bar\nu_{\alpha'}$ transition probability we need to make in (\ref{standard3}) the  change
$U_{\alpha i}\to U^{*}_{\alpha i}$. Thus, we have
\begin{equation}\label{standard4}
P(\bar \nu_\alpha\to \bar\nu_{\alpha'})=\delta _{\alpha' \alpha}
-4~\sum_{i>k}\mathrm{Re}~U_{\alpha'i}
U^{*}_{\alpha i}U^{*}_{\alpha'k}
U_{\alpha k}\sin^{2}\Delta _{ki}-2~\sum_{i>k}\mathrm{Im}~U_{\alpha'i}
U^{*}_{\alpha i}U^{*}_{\alpha'k}
U_{\alpha k}\sin 2\Delta _{ki}
\end{equation}
It is obvious that in the case of $\alpha'=\alpha$ the last terms of
(\ref{standard3}) and (\ref{standard4}) are equal to zero. We have
\begin{equation}\label{standard5}
P( \nu_\alpha\to \nu_{\alpha})=P(\bar \nu_\alpha\to \bar\nu_{\alpha}).
\end{equation}
This relation is a consequence of the $CPT$ invariance.

If $CP$ invariance in the lepton sector holds, in this case $U_{\alpha i}=U^{*}_{\alpha i}$ and
\begin{equation}\label{standard6}
P( \nu_\alpha\to \nu_{\alpha'})=P(\bar \nu_\alpha\to \bar\nu_{\alpha'})\quad \alpha'\neq \alpha.
\end{equation}
From (\ref{standard3}) and (\ref{standard4}) for the $CP$ asymmetry we have
\begin{equation}\label{CPasymmetry}
A^{CP}_{\alpha'\alpha}=P( \nu_\alpha\to \nu_{\alpha'})-P(\bar \nu_\alpha\to \bar\nu_{\alpha'})=4\sum_{i>k}A^{ik}_{\alpha'\alpha}~\sin2\Delta_{ki}.
\end{equation}
where
\begin{equation}\label{CPasymmetry1}
A^{ik}_{\alpha'\alpha}= \mathrm{Im}~U_{\alpha'i}
U^{*}_{\alpha i}U^{*}_{\alpha'k}
U_{\alpha k}.
\end{equation}
It obvious from (\ref{CPasymmetry1}) that
\begin{equation}\label{CPasymmetry2}
A^{ik}_{\alpha'\alpha}=-A^{ki}_{\alpha'\alpha}.
\end{equation}
In the case of the three-neutrino mixing the $CP$ asymmetry is characterized by $A^{21}_{\alpha'\alpha}$, $A^{31}_{\alpha'\alpha}$ and $A^{32}_{\alpha'\alpha}$. We will show now that (see \cite{Bilenky})
\begin{equation}\label{CPasymmetry3}
A^{21}_{\alpha'\alpha} =-A^{31}_{\alpha'\alpha}=A^{32}_{\alpha'\alpha}.
\end{equation}
In fact, taking into account the unitarity of the mixing matrix we have
\begin{equation}\label{CPasymmetry4}
\sum_{k}A^{ik}_{\alpha'\alpha}=0.
\end{equation}
From (\ref{CPasymmetry2}) and (\ref{CPasymmetry4}) we find
\begin{equation}\label{CPasymmetry5}
A^{12}_{\alpha'\alpha}+A^{13}_{\alpha'\alpha}=0,~~
A^{21}_{\alpha'\alpha}+A^{23}_{\alpha'\alpha}=0~~
A^{31}_{\alpha'\alpha}+A^{32}_{\alpha'\alpha}=0
\end{equation}
From these equations we easily find the relations (\ref{CPasymmetry3}).
Thus, in the three-neutrino case the $CP$ asymmetry has the form
\begin{equation}\label{CPasymmetry6}
A^{CP}_{\alpha'\alpha}=4~A^{21}_{\alpha'\alpha}~(\sin2\Delta_{12}
+\sin2\Delta_{23}-\sin2\Delta_{13})
\end{equation}
where
\begin{equation}\label{CPasymmetry7}
 \Delta_{13}=\Delta_{12} +\Delta_{23}.
\end{equation}
For any $a$ and $b$ we have the relation
\begin{equation}\label{CPasymmetry8}
\sin a+\sin b-\sin(a+b)-4\sin\frac{a+b}{2}\sin\frac{a}{2}\sin\frac{b}{2}.
\end{equation}
Thus, the $CP$ asymmetry in the three-neutrino case is given by the expression
\begin{equation}\label{CPasymmetry9}
A^{CP}_{\alpha'\alpha}=16~A^{21}_{\alpha'\alpha}~
\sin(\Delta_{12}+\Delta_{23})
\sin\Delta_{12}\sin\Delta_{23}.
\end{equation}

\section{Alternative way of the calculation of the transition probability}
We will present here a simple method of the calculation of the probability of the neutrino transition in vacuum. In expression
for the vacuum transition probability, presented below, the unitarity of the mixing matrix will be fully utilized. In particular, the  expression (\ref{CPasymmetry9}) for the CP asymmetry will be obtained directly from the general expression for the transition probability without any additional calculations.

Let us return back to the expression (\ref{probability2}). Taking into account the unitarity of the mixing matrix we can rewrite this expression in the form
\begin{eqnarray}\label{Genexp1}
  P(\nu_{\alpha}\to \nu_{\alpha'})&=&|\delta_{\alpha'\alpha}+
\sum_{i}U_{\alpha'i}~(e^{-2i\Delta _{pi}}-1)~
U^{*}_{\alpha i}|^{2}\nonumber\\&=&|\delta_{\alpha'\alpha}-2i\sum_{i}U_{\alpha'i}~
U^{*}_{\alpha i}e^{-i\Delta_{pi}}\sin\Delta_{pi}|^{2}
\end{eqnarray}
It is obvious that  the index $i$ in (\ref{Genexp1}) runs over values $i\neq p$.

From (\ref{Genexp1}) we have
\begin{eqnarray}
P(\nu_{\alpha}\to \nu_{\alpha'})&=&\delta_{\alpha'\alpha}
-4\sum_{i}|U_{\alpha i}|^{2}\sin^{2}\Delta_{pi}\delta_{\alpha'\alpha}
+4\sum_{i}|U_{\alpha' i}|^{2}|U_{\alpha' i}|^{2}\sin^{2}\Delta_{pi}
\nonumber\\
&+&8~\mathrm{Re}~\sum_{i>k}~U_{\alpha' i}U^{*}_{\alpha i}U_{\alpha'
k}^{*}U_{\alpha k}~e^{-i(\Delta_{pi}-\Delta_{pk})}\sin\Delta_{pi}\sin\Delta_{pk}
\label{Genexp3}
\end{eqnarray}
Finally, we find the following general expression for $\nu_{\alpha}\to \nu_{\alpha'}$ ($\bar\nu_{\alpha}\to \bar\nu_{\alpha'}$) transition probability
\begin{eqnarray}
P(\nua{\alpha}\to \nua{\alpha'})
=\delta_{\alpha' \alpha }
-4\sum_{i}|U_{\alpha i}|^{2}(\delta_{\alpha' \alpha } - |U_{\alpha' i}|^{2})\sin^{2}\Delta_{pi}\nonumber\\
+8~\sum_{i>k}\mathrm{Re}~U_{\alpha' i}U^{*}_{\alpha i}U^{*}_{\alpha'
k}U_{\alpha k}\cos(\Delta_{pi}-\Delta_{pk})\sin\Delta_{pi}\sin\Delta_{pk}\nonumber\\
\pm 8~\sum_{i>k}\mathrm{Im}~U_{\alpha' i}U^{*}_{\alpha i}U^{*}_{\alpha'
k}U_{\alpha k}\sin(\Delta_{pi}-\Delta_{pk})\sin\Delta_{pi}\sin\Delta_{pk}
\label{Genexp4}
\end{eqnarray}

\section{Three-neutrino oscillations}
\subsection{General expressions for $\nu_{l}\to \nu_{l'}$ ($\bar\nu_{l}\to \bar\nu_{l'}$) transition probabilities}

In the case of the three-neutrino mixing there are two independent mass-squared differences. From analysis of neutrino oscillation data it follows that one  mass-squared difference is much smaller than the other one. Correspondingly, two three-neutrino  mass spectra are possible
\begin{enumerate}
  \item Normal hierarchy (NH)
\begin{equation}\label{NS}
 m_{1}<m_{2}<m_{3},\quad \Delta m^{2}_{12}\ll \Delta m^{2}_{23}.
\end{equation}
\item Inverted hierarchy (IH)\footnote{Notice that neutrino masses are labeled differently in the case of NH and IH. This allows to introduce the same mixing angles in both cases}

\begin{equation}\label{IS}
m_{3}<m_{1}<m_{2},\quad \Delta m^{2}_{12}\ll |\Delta m^{2}_{13}|.
\end{equation}
\end{enumerate}
Let us denote two independent neutrino mass-squared differences $\Delta m^{2}_{S}$ (solar) and $\Delta m^{2}_{A}$ (atmospheric). We have
\begin{equation}\label{masssquared}
\Delta
m^{2}_{12}=\Delta
m^{2}_{S},~~m^{2}_{23}=\Delta
m^{2}_{A}~~(NH)\quad \Delta
m^{2}_{12}=\Delta
m^{2}_{S},~~\Delta m^{2}_{13}=-\Delta
m^{2}_{A}~~(IH).
\end{equation}
In the case of the NH it is natural to choose $p=2$. From the general expression (\ref{Genexp4}) we have
\begin{eqnarray}
&&P^{NH}(\nua{l}\to \nua{l'})
=\delta_{l' l }
-4\sum_{i}|U_{l 1}|^{2}(\delta_{l' l} - |U_{l' 1}|^{2})\sin^{2}\Delta_{S}\nonumber\\&&-4\sum_{i}|U_{l 3}|^{2}(\delta_{l' l} - |U_{l' 3}|^{2})\sin^{2}\Delta_{A}
-8~\mathrm{Re}~U_{l' 3}U^{*}_{l 3}U^{*}_{l'
1}U_{l 1}\cos(\Delta_{A}+\Delta_{S})\sin\Delta_{A}\sin\Delta_{S}\nonumber\\
&&\mp 8~\mathrm{Im}~U_{l' 3}U^{*}_{l 3}U^{*}_{l'
1}U_{l 1}\sin(\Delta_{A}+\Delta_{S})\sin\Delta_{A}\sin\Delta_{S}
\label{Genexp5}
\end{eqnarray}
In the case of the IH we choose $p=1$. For the transition probability
we obtain the following expression
\begin{eqnarray}
&&P^{IH}(\nua{l}\to \nua{l'})
=\delta_{l' l }
-4\sum_{i}|U_{l 2}|^{2}(\delta_{l' l } - |U_{l' 2}|^{2})\sin^{2}\Delta_{S}\nonumber\\&&-4\sum_{i}|U_{l 3}|^{2}(\delta_{l' l} - |U_{l' 3}|^{2})\sin^{2}\Delta_{A}
-8~\mathrm{Re}~U_{l' 3}U^{*}_{l 3}U^{*}_{l'
2}U_{l 2}\cos(\Delta_{A}+\Delta_{S})\sin\Delta_{A}\sin\Delta_{S}\nonumber\\
&&\pm 8~\sum_{i>k}\mathrm{Im}~U_{l' 3}U^{*}_{l 3}U^{*}_{l'
2}U_{l 2}\sin(\Delta_{A}+\Delta_{S})\sin\Delta_{A}\sin\Delta_{S}
\label{Genexp6}
\end{eqnarray}
The expressions (\ref{Genexp5}) and (\ref{Genexp6}) differ
by the change $U_{l 1}\to U_{l 2}$ and by the sign of the last term.
Notice that for the CP asymmetry directly from (\ref{Genexp5}) and (\ref{Genexp6}) we have
\begin{equation}\label{CPNH}
A^{CP}_{l' l }=-16~\mathrm{Im}~U_{l' 3}U^{*}_{l 3}U^{*}_{l'
1}U_{l 1}\sin(\Delta_{A}+\Delta_{S})\sin\Delta_{A}\sin\Delta_{S}
\end{equation}
in the case of NH and
\begin{equation}\label{CPIH}
A^{CP}_{l' l }=16~\mathrm{Im}~U_{l' 3}U^{*}_{l 3}U^{*}_{l'
1}U_{l 1}\sin(\Delta_{A}+\Delta_{S})\sin\Delta_{A}\sin\Delta_{S}
\end{equation}
in the case of IH.
In the standard parameterizations the $3\times3$ PMNS \cite{BPonte},\cite{MNS} mixing matrix $U$ is characterized by three mixing angles and one $CP$ phase and has the
form
\begin{eqnarray}
U=\left(\begin{array}{ccc}c_{13}c_{12}&c_{13}s_{12}&s_{13}e^{-i\delta}\\
-c_{23}s_{12}-s_{23}c_{12}s_{13}e^{i\delta}&
c_{23}c_{12}-s_{23}s_{12}s_{13}e^{i\delta}&c_{13}s_{23}\\
s_{23}s_{12}-c_{23}c_{12}s_{13}e^{i\delta}&
-s_{23}c_{12}-c_{23}s_{12}s_{13}e^{i\delta}&c_{13}c_{23}
\end{array}\right).
\label{unitmixU1}
\end{eqnarray}
Here $c_{12}=\cos\theta_{12}$,  $s_{12}=\sin\theta_{12}$ etc.

\subsection{Leading approximation}
From analysis of the neutrino oscillation data it follows that two neutrino oscillation parameters are small:
\begin{equation}\label{leading1}
 \frac{\Delta m^{2}_{S}}{\Delta m^{2}_{A}}\simeq 3\cdot 10^{-2},\quad
\sin^{2}\theta_{13}\simeq 2.4\cdot 10^{-2}
\end{equation}
In atmospheric region  of the parameter $\frac{L}{E}$ ($\frac{\Delta
m^{2}_{A}L}{2E}\gtrsim 1$) effects of neutrino oscillations are
large. In the first, leading  approximation we can neglect small contributions of
$\Delta m^{2}_{S}$ and $\sin^{2}\theta_{13}$ into neutrino transition
probabilities. From (\ref{Genexp5}), (\ref{Genexp6})  and (\ref{unitmixU1})
for the probability of $\nu_{\mu}$ ($\bar\nu_{\mu}$) to survive (for
both neutrino mass spectra) we obtain the following expression
\begin{eqnarray}\label{leading1}
P^{NH}(\nua{l}\to \nua{l'})&\simeq& P^{IH}(\nua{l}\to \nua{l'})
\simeq 1- 4 |U_{\mu 3}|^{2}(1-|U_{\mu 3}|^{2})\sin^{2}\Delta
m^{2}_{A}\frac{L}{4E}\nonumber\\&=&1-\sin^{2}2\theta_{23}\sin^{2}\Delta
m^{2}_{A}\frac{L}{4E}.
\end{eqnarray}
In the leading approximation we have $P(\nu_{\mu}\to \nu_{e})\simeq 0$ and
\begin{equation}\label{leading4}
P(\nu_{\mu}\to \nu_{\tau})\simeq 1-P(\nu_{\mu}\to \nu_{\mu})\simeq
\sin^{2}2\theta_{23}\sin^{2}\Delta
m^{2}_{A}\frac{L}{4E}.
\end{equation}
Thus, in the atmospheric region predominantly two-neutrino
$\nu_{\mu}\rightleftarrows \nu_{\tau}$ oscillations take place.

Let us consider now $\bar\nu_{e}\to \bar\nu_{e}$ transition in the
reactor Kamland region ($\frac{\Delta m^{2}_{S}L}{2E}\gtrsim 1$)   .
Neglecting contribution of $\sin^{2}\theta_{13}$ we have
\begin{equation}\label{leading2}
P^{NH}(\bar\nu_{e}\to \bar\nu_{e})\simeq P^{IH}(\bar\nu_{e}\to \bar\nu_{e})\simeq
1-\sin^{2}2\theta_{12}\sin^{2}\Delta
m^{2}_{S}\frac{L}{4E}.
\end{equation}
For appearance probabilities we find
\begin{equation}\label{leading6}
P^{NH}(\bar\nu_{e}\to \bar\nu_{\mu})\simeq
P^{IH}(\bar\nu_{e}\to \bar\nu_{\mu})\simeq
\sin^{2}2\theta_{12} \cos^{2}\theta_{23}
\sin^{2}\Delta
m^{2}_{S}\frac{L}{4E}
\end{equation}
and
\begin{equation}\label{leading7}
P^{NH}(\bar\nu_{e}\to \bar\nu_{\tau})\simeq
P^{IH}(\bar\nu_{e}\to \bar\nu_{\tau})\simeq
\sin^{2}2\theta_{12} \sin^{2}\theta_{23}
\sin^{2}\Delta
m^{2}_{S}\frac{L}{4E}
\end{equation}
We have
\begin{equation}\label{leading8}
P(\bar\nu_{e}\to \bar\nu_{e})= 1-P(\bar\nu_{e}\to \bar\nu_{\mu})
-P(\bar\nu_{e}\to \bar\nu_{\tau})
\end{equation}
and
\begin{equation}\label{leading9}
  \frac{P(\bar\nu_{e}\to
\bar\nu_{\tau})}{P(\bar\nu_{e}\to \bar\nu_{\mu})}\simeq
\tan^{2}\theta_{23}\simeq 1.
\end{equation}
Thus, in the reactor Kamland region $\bar\nu_{e}\rightleftarrows
\bar\nu_{\mu}$ and $\bar\nu_{e}\rightleftarrows \nu_{\tau}$
oscillations take place.

The expressions (\ref{leading1}) and (\ref{leading1}) were used for analysis of the first Super-Kamiokande atmospheric data, K2K and MINOS accelerator data and data of the reactor KamLAND experiment. Now with improved accuracy of the neutrino oscillation experiments  it is more common to perform more complicated three-neutrino analysis of the data.

\subsubsection{$\bar\nu_{e}\to \bar\nu_{e}$ survival probability}
From (\ref{Genexp5}) and  (\ref{Genexp6}) we can easily obtain exact three-neutrino expressions for $\bar\nu_{e}\to \bar\nu_{e}$
survival probabilities for both neutrino mass spectra. We have, correspondingly,
\begin{eqnarray}
&&P^{\mathrm{NH}}(\bar\nu_{e}\to \bar\nu_{e})=1- 4~|U_{e
3}|^{2}(1-|U_{e3}|^{2})~ \sin^{2}\frac{\Delta m^{2}_{A}L}{4E}
\nonumber\\
&&- 4~|U_{e 1}|^{2}(1-|U_{e 1}|^{2})~ \sin^{2}\frac{\Delta m^{2}_{S}L}{4E}
\nonumber\\
&&-8~|U_{e 3}|^{2}|U_{e
1}|^{2}~\cos\frac{(\Delta
m^{2}_{A}+\Delta m^{2}_{S})L}{4E}
\sin\frac{\Delta m^{2}_{A}L}{4E}\sin\frac{\Delta m^{2}_{S}L}{4E}.\label{3nue1}
\end{eqnarray}
and
\begin{eqnarray}
&&P^{\mathrm{IH}}(\bar\nu_{e}\to \bar\nu_{e})=1- 4~|U_{e
3}|^{2}(1-|U_{e3}|^{2})~ \sin^{2}\frac{\Delta m^{2}_{A}L}{4E}
\nonumber\\
&&- 4~|U_{e 2}|^{2}(1-|U_{e 2}|^{2})~ \sin^{2}\frac{\Delta m^{2}_{S}L}{4E}
\nonumber\\
&&-8~|U_{e 3}|^{2}|U_{e
2}|^{2}~\cos\frac{(\Delta
m^{2}_{A}+\Delta m^{2}_{S})L}{4E}
\sin\frac{\Delta m^{2}_{A}L}{4E}\sin\frac{\Delta m^{2}_{S}L}{4E}.\label{3nue2}
\end{eqnarray}
In the standard parameterization  of the PMNS mixing matrix we have
\begin{eqnarray}
&&P^{\mathrm{NS}}(\bar\nu_{e}\to \bar\nu_{e})=1-
\sin^{2}2\theta_{13} \sin^{2}\frac{\Delta m^{2}_{A}L}{2E}
\nonumber\\
&&-
(\cos^{2}\theta_{13}\sin^{2}2\theta_{12}+\sin^{2}2\theta_{13}\cos^{4}
\theta_{12}) ~ \sin^{2}\frac{\Delta m^{2}_{S}L}{2E}
\nonumber\\
&&-2\sin^{2}2\theta_{13}\cos^{2}\theta_{12} ~\cos\frac{(\Delta
m^{2}_{A}+\Delta m^{2}_{S})L}{4E} \sin\frac{\Delta
m^{2}_{A}L}{4E}\sin\frac{\Delta m^{2}_{S}L}{4E}.\label{3nue4}
\end{eqnarray}
and
\begin{eqnarray}
&&P^{\mathrm{IS}}(\bar\nu_{e}\to \bar\nu_{e})=1-
\sin^{2}2\theta_{13} \sin^{2}\frac{\Delta m^{2}_{A}L}{2E}
\nonumber\\
&&-
(\cos^{2}\theta_{13}\sin^{2}2\theta_{12}+\sin^{2}2    \theta_{13}\sin^{4}
\theta_{12}) ~ \sin^{2}\frac{\Delta m^{2}_{S}L}{2E}
\nonumber\\
&&-2\sin^{2}2\theta_{13}\sin^{2}\theta_{12} ~\cos\frac{(\Delta
m^{2}_{A}+\Delta m^{2}_{S})L}{4E} \sin\frac{\Delta
m^{2}_{A}L}{4E}\sin\frac{\Delta m^{2}_{S}L}{4E}.\label{3nue5}
\end{eqnarray}

\subsubsection{$\nu_{\mu}\to \nu_{e}$ ($\bar\nu_{\mu}\to \bar\nu_{e}$)
 transition probabilities}
From (\ref{Genexp5}) and (\ref{Genexp6}) we  obtain the following expressions for $\nua{\mu}\to \nua{e}$ vacuum transition probabilities:
\begin{eqnarray}
&&P^{\mathrm{NS}}(\nua{\mu}\to \nua{e})= 4~|U_{e 3}|^{2}|U_{\mu
3}|^{2}~ \sin^{2}\Delta_{A}
\nonumber\\
&&+4~|U_{e 1}|^{2}|U_{\mu 1}|^{2}~
\sin^{2}\Delta_{S}\nonumber\\
&&-8~\mathrm{Re}~U_{e 3}U^{*}_{\mu3}U_{e 1}^{*}U_{\mu 1}~\cos(\Delta
_{A}+\Delta _{S}) \sin\Delta
_{A}\sin\Delta _{S}\nonumber\\
&&\mp 8~\mathrm{Im}~U_{e 3}U^{*}_{\mu 3}U_{e 1}^{*}U_{\mu
1}~\sin(\Delta _{A}+\Delta _{S}) \sin\Delta _{A}\sin\Delta _{S}
.\label{3numue1}
\end{eqnarray}
and
\begin{eqnarray}
&&P^{\mathrm{IS}}(\nua{\mu}\to \nua{e})= 4~|U_{e 3}|^{2}|U_{\mu
3}|^{2}~ \sin^{2}\Delta_{A}
\nonumber\\
&&+4~|U_{e 2}|^{2}|U_{\mu 2}|^{2}~
\sin^{2}\Delta_{S}\nonumber\\
&&-8~\mathrm{Re}~U_{e 3}U^{*}_{\mu3}U_{e 2}^{*}U_{\mu 2}~\cos(\Delta
_{A}+\Delta _{S}) \sin\Delta
_{A}\sin\Delta _{S}\nonumber\\
&&\pm 8~\mathrm{Im}~U_{e 3}U^{*}_{\mu 3}U_{e 2}^{*}U_{\mu
2}~\sin(\Delta _{A}+\Delta _{S}) \sin\Delta _{A}\sin\Delta _{S}
.\label{3numue2}
\end{eqnarray}
Using the standard parameterization of the PMNS mixing
matrix in the case of NH we have
\begin{eqnarray}
&&P^{\mathrm{NS}}(\nua{\mu}\to
\nua{e})=\sin^{2}2\theta_{13}s^{2}_{23}\sin^{2}\frac{\Delta
m^{2}_{A}L}{4E}\nonumber\\&&+(\sin^{2}2\theta_{12}c^{2}_{13}c^{2}_{23}+
\sin^{2}2\theta_{13}c^{4}_{12}s^{2}_{23}+Kc^{2}_{12}\cos\delta)
\sin^{2}\frac{\Delta
m^{2}_{S}L}{4E}\nonumber\\&&+(2\sin^{2}2\theta_{13}s^{2}_{23}c^{2}_{12}+K\cos\delta)
~\cos\frac{(\Delta m^{2}_{A}+\Delta m^{2}_{S})L}{4E} \sin\frac{\Delta
m^{2}_{A}L}{4E}\sin\frac{\Delta m^{2}_{S}L}{4E}\nonumber\\
&&\mp K\sin\delta~~\sin\frac{(\Delta m^{2}_{A}+\Delta
m^{2}_{S})L}{4E} \sin\frac{\Delta m^{2}_{A}L}{4E}\sin\frac{\Delta
m^{2}_{S}L}{4E}. \label{3numue3}
\end{eqnarray}
Here
\begin{equation}\label{3numue4}
  K=\sin2\theta_{12}\sin2\theta_{13}\sin2\theta_{23}c_{13}.
\end{equation}
In the case of the inverted neutrino mass hiearchy we find the
following expressions for the transition probabilities
\begin{eqnarray}
&&P^{\mathrm{IS}}(\nua{\mu}\to
\nua{e})=\sin^{2}2\theta_{13}s^{2}_{23}\sin^{2}\frac{\Delta
m^{2}_{A}L}{4E}\nonumber\\&&+(\sin^{2}2\theta_{12}c^{2}_{13}c^{2}_{23}+
\sin^{2}2\theta_{13}s^{4}_{12}s^{2}_{23}-Ks^{2}_{12}\cos\delta)
\sin^{2}\frac{\Delta
m^{2}_{S}L}{4E}\nonumber\\&&+(2\sin^{2}2\theta_{13}s^{2}_{23}s^{2}_{12}-K\cos\delta)
~\cos\frac{(\Delta m^{2}_{A}+\Delta m^{2}_{S})L}{4E} \sin\frac{\Delta
m^{2}_{A}L}{4E}\sin\frac{\Delta m^{2}_{S}L}{4E}\nonumber\\
&&\mp K\sin\delta~~\sin\frac{(\Delta m^{2}_{A}+\Delta
m^{2}_{S})L}{4E} \sin\frac{\Delta m^{2}_{A}L}{4E}\sin\frac{\Delta
m^{2}_{S}L}{4E}. \label{3numue5}
\end{eqnarray}
Formulas (\ref{3numue3}) and (\ref{3numue5}) can be used for analysis of the data of T2K long baseline accelerator  experiment in which matter effects are negligible.

\section{Transitions in the case of sterile neutrinos}
From  existing data some indications in favor of
neutrino oscillations driven by "large"
($\sim 1 \mathrm{eV}^{2}$) neutrino mass-squared difference(s)
were obtained.(see \cite{Carlo}, \cite{Whitepap}). These data (if correct) would proof existence of sterile neutrinos.

From general formula (\ref{Genexp4}) we can easily obtain transition probabilities in such cases. In the framework of mixing of  four massive neutrinos we will consider first the simplest 3+1 scheme in which the forth mass is separated from three close masses by a $\sim 1 \mathrm{eV}$ gap.
Let us choose $p=1$. In the region of $\frac{L}{E}$ sensitive to large
neutrino mass-squared difference  ($\frac{\Delta m^{2}L}{4E}\gtrsim 1$,~~~$\Delta m^{2}\equiv\Delta m_{14}^{2}$)  we have
$\Delta_{1i}\simeq 0,~~i=2,3$. From  (\ref{Genexp4}) we find in this case
\begin{equation}\label{ster}
 P(\nua{\alpha}\to \nua{\alpha'})
\simeq \delta_{\alpha' \alpha }
-4|U_{\alpha 4}|^{2}(\delta_{\alpha' \alpha } - |U_{\alpha' 4}|^{2})\sin^{2}\frac{\Delta m^{2}L}{4E}.
\end{equation}
In the case of reactor antineutrinos we have
\begin{equation}\label{ster1}
 P(\bar\nu_{e}\to \bar\nu_{e})
=1
-4~|U_{e 4}|^{2}(1- |U_{e 4}|^{2})\sin^{2}\frac{\Delta m^{2}L}{4E}
\end{equation}
For $\bar\nu_{\mu}\to \bar\nu_{e}$ transition (LSND) we find
\begin{equation}\label{ster2}
 P(\bar\nu_{\mu}\to \bar\nu_{e})
= 4~|U_{e 4}|^{2} |U_{\mu 4}|^{2})\sin^{2}\frac{\Delta m^{2}L}{4E}.
\end{equation}
Let us consider now  3+2 scheme with 2 masses $m_{4}$  and $m_{5}$ separated from three close masses by  a $\sim 1 \mathrm{eV}$ gaps.
Let us choose $p=1$. In the region of $\frac{L}{E}$ sensitive to
$\Delta m^{2}_{14}$ and $\Delta m^{2}_{15}$ we have $\Delta_{1i}\simeq 0,~~i=2,3$. From (\ref{Genexp4}) we find in this case
\begin{eqnarray}
P(\nua{\mu}\to \nua{e})
= 4|U_{e 4}|^{2} |U_{\mu 4}|^{2})\sin^{2}\frac{\Delta m_{14}^{2}L}{4E}
+4|U_{e 5}|^{2} |U_{\mu 5}|^{2})\sin^{2}\frac{\Delta m_{15}^{2}L}{4E}
\nonumber\\
+8~\mathrm{Re}~U_{e 5}U^{*}_{\mu 5}U^{*}_{e
4}U_{\mu 4}\cos(\frac{\Delta m_{15}^{2}L}{4E}-\frac{\Delta m_{14}^{2}L}{4E})\sin\frac{\Delta m_{15}^{2}L}{4E}\sin\frac{\Delta m_{14}^{2}L}{4E}\nonumber\\
\pm 8~\mathrm{Im}~U_{e 5}U^{*}_{\mu 5}U^{*}_{e
4}U_{\mu 4}\sin(\frac{\Delta m_{15}^{2}L}{4E}-\frac{\Delta m_{14}^{2}L}{4E})\sin\frac{\Delta m_{15}^{2}L}{4E}\sin\frac{\Delta m_{14}^{2}L}{4E}
\label{Ster4}
\end{eqnarray}
For $\nua{\alpha}\to \nua{\alpha}$ survival probability we find the following expression

\begin{eqnarray}
P(\nua{\alpha}\to \nua{\alpha'})
&=&1
-4|U_{\alpha 4}|^{2}(1 - |U_{\alpha 4}|^{2})\sin\frac{\Delta m_{14}^{2}L}{4E}-4|U_{\alpha 5}|^{2}(1 - |U_{\alpha 5}|^{2})\sin\frac{\Delta m_{15}^{2}L}{4E}\nonumber\\
+&&8~U_{\alpha 5}|^{2}|U_{\alpha
4}|^{2}\cos(\frac{\Delta m_{15}^{2}L}{4E}-\frac{\Delta m_{14}^{2}L}{4E})\sin\frac{\Delta m_{15}^{2}L}{4E}\sin\frac{\Delta m_{14}^{2}L}{4E}
\label{Ster5}
\end{eqnarray}

I am thankful to A. Olshevskiy and C. Giunti for useful discussion.

\end{document}